\documentstyle[11pt,newpasp,twoside,epsf]{article}
\newcommand{\smyr}{\ M_\odot\ {\rm yr^{-1}}}
\newcommand{\sm}{\ M_\odot}
\newcommand{\kms}{\ {\rm km\:s^{-1}}}
\def\lesssim{\mathrel{\hbox{\rlap{\hbox{\lower4pt\hbox{$\sim$}}}\hbox{$<$}}}}
\def\gtrsim{\mathrel{\hbox{\rlap{\hbox{\lower4pt\hbox{$\sim$}}}\hbox{$>$}}}}
\newcommand{\cth}{c_{\rm th}}

\newcommand{\beq}{\begin{equation}}
\newcommand{\eeq}{\end{equation}}

\newcommand{\ecore}{\epsilon_{\rm core}}

\def\ion#1#2{#1$\;${\small\rm II}\relax}

\def\edcomment#1{\iffalse\marginpar{\raggedright\sl#1\/}\else\relax\fi}
\reversemarginpar

\begin{document}
\title{Theories of Massive Star Formation: Collisions, Accretion and the View from the ``I'' of Orion}
 \author{Jonathan C. Tan}
\affil{Princeton University Observatory, Princeton, NJ 08544, USA}

\begin{abstract}
I review the arguments motivating models for massive star formation
via stellar collisions. I then describe how
the standard accretion scenario, involving the collapse of a
quasi-hydrostatic gas core, can produce high-mass stars in the
pressurized regions of forming star clusters. I argue that the
observational evidence, particularly in the Orion hot core, favors the
standard accretion paradigm.
\end{abstract}

\section{Introduction}


Two basic models for how a massive star accumulates its mass are
debated. The conventional case, which I refer to as the {\it standard
accretion model}, involves the inside-out collapse of a
gravitationally bound, pre-stellar gas core from approximate
hydrostatic equilibrium. A small amount of angular momentum creates an
accretion disk, via which mass reaches the star in an ordered
manner. Low-mass stars appear to form in this way (Shu, Adams, \&
Lizano 1987).

The {\it collisional model} involves the coalescence of smaller stars
or protostars in dense stellar clusters (Bonnell, Bate, \& Zinnecker
1998). These lower-mass stars may first form via standard accretion from
small gas cores that have fragmented from the protocluster gas
clump. They may also build up their mass via Bondi-Hoyle accretion
from the clump.

\section{The Case Against Standard Accretion}

Some of these points
are discussed in the review by Stahler, Palla, \&
Ho (2000).

{\bf Long Formation Time:} Assuming 100\% efficiency, the accretion
rate of a star forming from the collapse of a critically unstable
isothermal sphere is $\dot{m}_*=0.975\cth^3/G=4.4\times 10^{-6}
(T/20\:{\rm K})^{3/2}\smyr$ (Shu 1977), so the formation time is
$t_{*f}=m_{*f}/\dot{m}_*=6.9\times 10^6 (T/20\:{\rm K})^{-3/2}
(m_{*f}/30\sm)\:{\rm yr}$. This timescale should be less than the main
sequence lifetime (several Myr for massive stars) and the cluster
formation time ($\sim 1$~Myr for the Orion Nebula Cluster (ONC); Palla
\& Stahler 1999), requiring large accretion rates and temperatures
(e.g. $\dot{m}_*>10^{-4}\smyr$ and $T>160\:{\rm K}$ for
$m_{*f}=100\sm$ and $t_{*f}<1$~Myr).  Such temperatures are observed
in massive star-forming regions, but are unlikely to occur until at
least one massive star is present. Higher accretion rates result if a
core has some nonthermal pressure support (Stahler, Shu, \& Taam
1980). This does appear to be the case for massive cores since their
line-widths correspond to supersonic velocities.

{\bf Radiation Pressure:} The short Kelvin-Helmholtz times of massive
stars allow them to reach the main sequence while they are still
accreting (e.g. Palla \& Stahler 1992). These protostars have high
luminosity, $L$, and the radiation pressure may disrupt infall by
acting on dust grains, which survive until temperatures of $\sim
2300\:{\rm K}$ and are well-coupled to the dense gas. A necessary
condition for infall is that the ram pressure of inward motion of the
gas exceed the outward radiation pressure from the direct radiation
field of the star at the dust destruction front, $r_d \simeq 13
(L/10^5{\rm L_\odot})^{1/2}(T_d/{\rm 2300K})^{-2}(Q_a/0.1)^{-1/2}{\rm
AU}$, where $Q_a$ is the effective absorption efficiency of dust
grains at the sublimation temperature. This implies $\rho v^2>L/(4\pi
r_d^2 c)$, where $\rho$ and $v$ are the density and infall velocity of
the gas at $r_d$. In the spherical case we then have $\dot{m}>L/(c
v)$. For zero age main sequence (ZAMS) stars of mass $30, 60, 120\sm$,
$L\simeq 1.2,5.1,18\times 10^5\:{\rm L_\odot}$ (Schaller et al. 1992),
and assuming free-fall conditions at $r_d$ we have $\dot{m}\gtrsim
0.4,1.7,6.2\times 10^{-4}\smyr$ (see also Wolfire \& Cassinelli
1987). Nakano (1989) and Jijina \& Adams (1996) have argued that these
conditions are relaxed for accretion via a disk, since once gas
reaches the disk (typically at radii $\gg r_d$), it is shielded from
much of the stellar radiation. However, the ram pressure criterion
should still apply in the disk at $r_d$. Consideration of this case
(Tan \& McKee 2002, in prep.) suggests that accretion rates $\sim 10^{-3}\smyr$
are required to form the most massive stars. Radiation-hydrodynamic
simulations (e.g. Yorke \& Sonnhalter 2002) will ultimately provide
the most accurate answer to this question. 

{\bf Crowding:} It appears that massive stars form almost exclusively in
clusters containing many more low-mass stars, which dominate the total
stellar mass. The central stellar densities can be very
high---Hillenbrand \& Hartmann (1998) find $n_*\simeq 1.7\times
10^{4}\:{\rm pc^{-3}}$ in the ONC corresponding to separations of
about 0.04 pc. The hydrostatic cores invoked in standard accretion
would have to form and survive in such a crowded environment.

{\bf Small Jeans Mass:} The Jeans mass (or more appropriately the
Bonnor-Ebert mass, which is the maximum mass of a stable isothermal
sphere) becomes very small in regions of high pressure, such as
high-mass star-forming clumps where $\bar{P}\sim
10^8-10^9\:{\rm K\:cm^{-3}}$ (McKee \& Tan 2002b [MT02b]). The Bonnor-Ebert
mass is $M_{\rm BE}=0.046(T/20\:{\rm K})^2 (\bar{P}/10^9\:{\rm
K\:cm^{-3}})^{-1/2}M_\odot$. If gas forms bound structures only on this
scale, then to form a massive star, most of the mass would have to be
accumulated via Bondi-Hoyle accretion or mergers.


\section{The Case for Collisions and Competitive Accretion}

A model of massive star formation by collisions of lower-mass stars
(Bonnell et al. 1998), whose masses may be augmented by
competitive Bondi-Hoyle accretion (Bonnell et al. 2001a), can
automatically circumvent three of the four difficulties faced by
standard accretion.
Firstly, the coalescence of stars, or any optically thick mass unit, is not
prevented by radiation pressure. However, Bondi-Hoyle accretion is
suppressed, i.e. for $m_*\gtrsim 10\sm$.  Rates of Bondi-Hoyle
accretion are enhanced in dense regions and collisions are favored in
crowded regions---just the locations where massive stars are observed
to form. Bonnell \& Davies (1998) have shown that there has not been
enough time for the Trapezium stars in the ONC to have reached their
central location by dynamical relaxation. Either they formed in the
center (half of the massive stars forming within 30\% of the half-mass
radius) or they formed with much smaller velocity dispersions than the
lower mass stars. Bondi-Hoyle accretion and mergers can potentially
build up sub-solar Bonnor-Ebert masses to the massive star
regime. A simulation of competitive accretion (Bonnell et
al. 2001b) has reproduced the observed stellar IMF from an initial
cluster of a 1000 identical $0.1\sm$ stars embedded in a cold gas
clump that makes up 90\% of the mass. Note, however, that in
some star-forming regions pre-stellar gas cores are observed to have a mass
spectrum similar to that of stars (up to at least $5\sm$) (e.g. Motte et al. 2001).

It has been proposed that run-away OB stars, which are ejected from
clusters with velocities $\sim 50-150\kms$, and the small primary to
secondary mass ratios of massive star binaries, may be the by-products
of massive star formation by collisions (Stahler et al. 2000). Both
phenomena are expected to result from close three body interactions of
massive stars---the least massive star tends to be ejected leaving the
remaining stars in a tightened binary. Such encounters should be
common if massive star mergers are occurring. An alternative mechanism
for producing run-away OB stars (from clusters that are at least
several Myr old) is a supernova in a double massive star binary
system. Hoogerwerf, de Bruijne, \& de Zeeuw (2001) have identified the
ejected stars from single examples of both of these types of event.

It is clear that collisions and strong dynamical interactions can
occur in the crowded environments of star clusters, but the real
question is whether the rate is great enough to be relevant for the star
formation process. The collisional timescale in a cluster of equal
mass stars in the strong gravitational focusing limit is (Binney \&
Tremaine 1987)
\beq
\label{tcoll}
t_{\rm coll}=1.44\times 10^{10} \left(\frac{10^4{\rm
pc^{-3}}}{n_*}\right)\left(\frac{\sigma}{2\kms}\right)\left(\frac{10{\rm
R_\odot}}{r_*}\right)\left(\frac{M_\odot}{m_*}\right)\:{\rm yr} \eeq
where $\sigma$ is the 1D velocity dispersion, and $r_*$ and $m_*$ the
stellar radius and mass. In the above equation we have normalized the
variables to values typical of the central region of the ONC and
allowed for a generous increase in the stellar radius due to pre-main
sequence activity. For collisions to be important we require the
collisional timescale to be $\lesssim 10^6\:{\rm yr}$, which requires
a substantial change in one or more of the above parameters.

Stahler et al. (2000) pointed out that the collisional cross section
is substantially increased for protostars that are still embedded in
dense gas cores or have massive disks (see also McDonald \& Clarke
1995). However, it is also possible that such encounters separate the
stars from their gas without leading to a collision (Price \&
Podsiadlowski 1995), which would reduce the efficiency of this
mechanism. Zwart et al. (2001) have found that mass segregation and
binarity can boost the collision rate. It may also be increased if
stars are swollen by the energy release from a recent collision.
Nevertheless, even for an optimistic collision radius of $1{\rm
AU}=215 {\rm R_\odot}$ (for each star) and for stellar masses
$m_*=10M_\odot$ equation (\ref{tcoll}) shows that densities of order
$10^6{\rm pc^{-3}}$ are needed for $t_{\rm coll}\sim 10^6{\rm
yr}$. More realistic cross sections suggest the densities need to be
$\gtrsim10^8{\rm pc^{-3}}$.

Bonnell et al. (1998) have presented a model in which extreme
densities can result in a cluster of lower-mass stars that are
accreting from their protocluster gas cloud.  The
velocity dispersion of the stars decreases as they accrete, lowering
the total energy of the system and causing the cluster to contract. In
a subsequent version of this model (Bonnell 2002) the initial gas
cloud, which dominates the total mass, is in free-fall collapse. The
cloud is seeded with various spatial distributions of low-mass
stars. Gas densities increase rapidly at the center and low-mass stars
that happen to be here grow quickly with large rates of Bondi-Hoyle
accretion (Bonnell et al 2001a,b). After one free-fall time the
central stellar density can increase from its initial value by factors
of about $10^5$. However, this dramatic increase appears to depend on
the assumption of free-fall collapse of the entire gas cloud. It is
argued that if the {\it initial} conditions for this simulation are
taken to be the {\it present-day} densities of the ONC, then
collisional massive star formation is likely to occur.

Finally, it has been suggested that filamentary structure in the
collapsing gas could provide a means of weakly collimating any
outflows that are produced in the star formation process (Bonnell
2002). It would be difficult for the collisional model to produce
coherent and symmetric outflows.

Two key observational predictions of the collisional model are the
sporadic release of large amounts of energy from stellar collisions
and the brief existence of dense ($\gtrsim 10^6-10^8{\rm pc^{-3}}$)
stellar clusters around each forming massive star.  Neither of these
have been observed, although they may be obscured by the high
extinction ($A_V\gtrsim 200$) typical of massive star-forming regions.

\section{Turbulent Core Accretion}

In spite of the difficulties facing the standard accretion
model (\S2), the presence of coherent and massive gas cores in
high-mass star-forming regions (Garay \& Lizano 1999; Kurtz et
al. 2000) suggests that a model for massive star formation
should start by considering the collapse of such structures
(McLaughlin \& Pudritz 1997; Osorio et al. 1999; McKee \& Tan 2002a [MT02a]; MT02b). Massive stars do not form in isolation and in the model of
MT02a,b we assume the {\it cores}, which may form
individual or binary stars, are part of a self-similar hierarchy of
structure. We term the gas cloud that forms a star cluster a {\it
clump}. The density distributions of clumps and cores, averaging over
internal clumpiness, are consistent with approximately spherically
symmetric power law density profiles, with $\rho\propto r^{-k_\rho}$
and $k_\rho\simeq 1.5$ (Evans et al. 2002).  Such structures can
result from the hydrostatic equilibrium of gas with an equation of
state $P\propto \rho^{\gamma_p}$, with $\gamma_p=2/3$.  Since the gas
is observed to be cooler than the virial temperature, it must be
supported by nonthermal forms of pressure. The signal speed,
$c\equiv(P/\rho)^{1/2}$, is therefore supersonic and so the cores and
the clumps should be turbulent and clumpy, as is observed. In the
fiducial case of $k_\rho=3/2$, $c\propto r^{1/4}$ and $P\propto
r^{-k_P}$, with $k_P=1$.  We define the surface of a core to be where
the pressure has decreased to the ambient pressure in the clump.  This
is set by self-gravity and is $\bar{P}_{\rm cl}\simeq 0.88G\Sigma_{\rm
cl}^2=8.5\times 10^8 \Sigma_{\rm cl}^2\:{\rm K\:cm^{-3}}$ in the
central regions where massive stars typically form. In this expression
$\Sigma_{\rm cl}\equiv M_{\rm cl}/(\pi R_{\rm cl}^2)$ is the mean
surface density of the clump, with characteristic value $\sim
1.0\:{\rm g\:cm^{-2}}$ (Plume et al. 1997).

Since the core mass inside a distance $r$ from its center is
$M(<r)=k_P c^2r/G$, then for higher ambient pressures (larger values
of $c^2$ at the core surface), the equilibrium state for a given core
mass becomes smaller and denser.  The efficiency ($\ecore\equiv
m_{*f}/M_{\rm core}$) of star formation from a core is typically quite
high for low-mass stars ($\sim 50\%$--Matzner \& McKee 2000), if set
by magneto-centrifugally driven bipolar outflows such as
disk-winds or X-winds (K\"onigl \& Pudritz 2000; Shu et al. 2000) that
launch a fraction $f_w\sim 0.1-0.3$ of the accreted mass with force
$\dot{p}_w= \dot{m}_w v_w = f_w \dot{m}_* v_w = \phi_w \dot{m}_* v_K$.
Here $v_K$ is the Keplerian velocity at the equatorial stellar radius
and $\phi_w=0.6$ is a fiducial value. These winds sweep up
and eject gas from polar regions of the core. Similarly high
efficiencies hold for massive stars for the same values of $f_w$ and
$\phi_w$ (Tan \& McKee 2002, in prep.). 
For $\ecore=0.5$ we have (MT02b)
\begin{equation}
\label{eq:rcore}
r_{\rm core}=0.057 \left(\frac{m_{*f}}{30\sm}\right)^{1/2}\Sigma_{\rm cl}^{-1/2}\:{\rm pc}.
\end{equation}
Recall the mean stellar separation in the center of the ONC of about
0.04 pc. Comparison with the expected size of a massive core shows
that crowding is not necessarily a problem, particularly since during
a typical stage of formation half the stars observed today were not
yet present, in the central part of the ONC there has been time for some relaxation and density
increase since formation, and most stars are
low-mass and have little dynamical impact on a massive core.

When a core does collapse it must be from a configuration at
least as dense as implied by equation (\ref{eq:rcore}). The high
pressures of massive star-forming clumps require dense cores, which
have short free-fall times. Inside-out collapse proceeds as
in the model of Shu (1977), but now, since $c\propto r^{1/4}$, the
accretion rate grows:
\begin{equation}
\label{mdotSigma}
\dot{m}_*= 4.6\times 10^{-4}
\left(\frac{m_{*f}}{30\:{\rm M_\odot}}\right)
^{3/4} \Sigma_{\rm cl}^{3/4} 
\left(\frac{m_*}{m_{*f}}\right)^{0.5}~{\rm M_\odot\:yr^{-1}},
\label{mdotSigma2}
\end{equation}
where $m_*$ is the instantaneous protostellar mass. It will fluctuate
because of core clumpiness. These mean accretion rates are large
enough to overcome the radiation pressure of massive stars (Tan \&
McKee 2002, in prep.). The corresponding formation timescales,
\begin{equation}
\label{tsfSigma}
t_{*f}= 1.29 \times 10^{5} 
\left(\frac{m_{*f}}{30\:{\rm M_\odot}}\right)^{1/4} 
\Sigma_{\rm cl}^{-3/4}~~~ {\rm yr}
\end{equation}
are short compared with the age constraints of the ONC
(Palla \& Stahler 1999).

In these models the mass of a star is
determined by the initial core mass and the efficiency of star
formation, $\ecore$. Since we find $\ecore$ to be relatively constant
with stellar mass, the shape of the stellar IMF should reflect that of
the cores. This is consistent with observations (e.g. Motte et
al. 2001). Such a mass function may result from the coagulation of
cores or from fluctuations in the turbulent velocity field that
produce unstable condensations, but we have not attempted to predict
it. We do not expect much hierarchical fragmentation of collapsing
cores if the collapse happens quite quickly after formation
(i.e. within a few dynamical times) and if most turbulent fluctuations
do not produce unstable cores. The central concentration of the cores
($\rho\propto r^{-1.5}$) may suppress fragmentation (Bodenheimer et
al. 2000). Once a protostar has formed, its tidal field will also
stabilize the collapsing core against fragmentation.




\begin{figure}  
\plotone{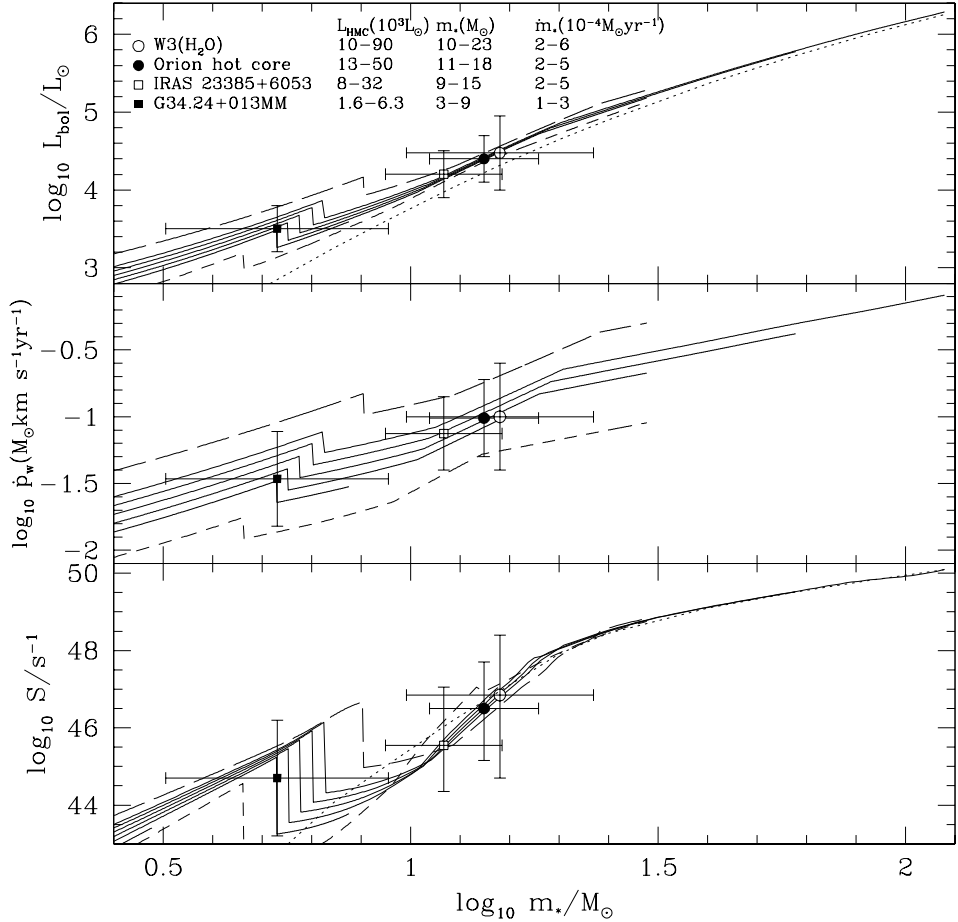}
\label{fig:jtan1}
\caption{Models of protostellar evolution: bolometric luminosity (top
panel), outflow force (middle panel), and ionizing luminosity (bottom
panel) versus protostellar mass. {\it Solid} lines show model
predictions for $m_{*f}=7.5,15,30,60,120\sm$ stars forming in a
$\Sigma_{\rm cl}=1{\rm g\:cm^{-2}}$ clump. {\it Dashed} and {\it long-dashed}
lines show a $30\sm$ star forming from $\Sigma_{\rm cl}=0.32,3.2{\rm g\:cm^{-2}}$
clumps respectively. The {\it dotted} line shows the ZAMS bolometric
(Schaller et al. 1992) and ionizing luminosities (Schaerer \& de Koter
1997; W. Vacca, private comm.). Constraints on the
properties of several nearby protostars are derived from the
luminosities of their hot molecular cores (see Osorio, Lizano, \&
D`Alessio 1999 for additional modeling and observational references).
}
\end{figure}

We have calculated the evolution of protostellar radius and luminosity
(bolometric, outflow-mechanical, and ionizing) for the accretion rates
predicted from the turbulent core collapse model. Before reaching the
ZAMS, the radius is determined by deuterium core and shell burning
(e.g. Palla \& Stahler 1992). Our model is based on that of Nakano et
al. (1995), but extended to include shell burning with a simple
prescription that matches the results of Palla \& Stahler.  With the
protostellar radius and accretion rate we calculate the accretion
luminosity, which adds to the internal luminosity to give the
total. The outflow force, $\dot{p}_w$, is calculated assuming
$\phi_w=0.6$, as described above. The rate of emission of H ionizing
photons, $S$, is calculated with contributions from the star and the
accretion shock and utilizing models of Schaerer \& de Koter (1997)
for solar metallicity. The models are shown in
Fig. 1. From the observed bolometric luminosities of
hot molecular cores, we constrain masses and accretion rates of
embedded protostars, assuming $\Sigma_{\rm cl}=1{\rm
g\:cm^{-2}}$. This also constrains $\dot{p}_w$ and $S$.

If the mechanism of magneto-centrifugal outflow generation remains
qualitatively similar to that operating in lower-mass stars, where
theoretical models (Shu et al. 2000; K\"onigl \& Pudritz 2000) predict
$f_w=\dot{m}_w/\dot{m}_*\sim 0.1-0.3$ with most of the outflow
originating from the inner ($r\lesssim 10r_*$) accretion disk, then
the large accretion rates predicted by equation (\ref{mdotSigma})
imply a high density of the outflowing gas rising from the disk.  For
fiducial parameters, we find that the ionizing fluxes of protostars at
least as massive as $\sim 60M_\odot$ are confined by their outflows in
directions approximately in the plane of the accretion disk.
Thus we do not expect disk photoevaporation (Hollenbach et al. 1994)
to be an important process during the accretion phase of most massive
stars.

Ionizing photons can more readily escape along the polar directions of
the flow, where MHD-outflow models predict the existence of
``dead-zone'' cavities that the outflow does not penetrate. These
regions should be occupied by a more diffuse conventional stellar
wind. The angular extent of the cavity as viewed from the star
decreases with distance as the outflow is collimated by
the hoop stresses of the toroidal magnetic field component. Outside
the cavity, the density of the outflow may be derived given the
angular distribution of the outflow force and assuming a constant
terminal velocity. For the force distribution of Matzner \&
McKee (1999), which includes the small angle softening parameter
$\theta_0$, we have
\begin{equation}
n_w = \frac{4.34\times10^5}{\rm sin^2\theta +\theta_0^2} \left(\frac{f_w}{0.1}\frac{10{\rm AU}}{r}\right)^{2} \frac{0.6}{\phi_w}\frac{\rm ln(200)}{\rm ln(2/\theta_0)}\frac{\dot{m}_*}{10^{-4}{M_\odot/{\rm yr}}} \left(\frac{30{\rm M_\odot}}{m_*}\frac{r_*}{\rm 10R_\odot}\right)^{1/2}{\rm cm^{-3}}.\label{nangular}
\end{equation}
A test of this model is the presence of very compact
\ion{H}{2} regions, embedded in the outflows of massive
protostars. These regions will initially be elongated along the
outflow axis, but should eventually ionize the entire flow as the star's
luminosity grows and the accretion and outflow rates finally
diminish. This transition should be quite rapid for a flow
density distribution declining as $(r {\rm sin}\theta)^{-2}$.

Of the hot cores in Fig. 1, 
No centimeter emission is seen (to
$\sim0.5$~mJy) from IRAS~23385+6053 and G34.24+0.13MM, perhaps because
of their lower luminosities and further distances. The Orion hot core
contains a very compact thermal source (``I''), discussed
below. W3($\rm H_2O$) contains a narrow radio jet (Wilner et
al. 1999). Shepherd \& Kurtz (1999) and Zhang et al. (2002) report
radio and SiO jets aligned with outflows in G192.16-3.82 and AFGL
5142. See also Garay \& Lizano (1999).

\section{The Orion Hot Core and Source ``I''}

\begin{figure}  
\plotfiddle{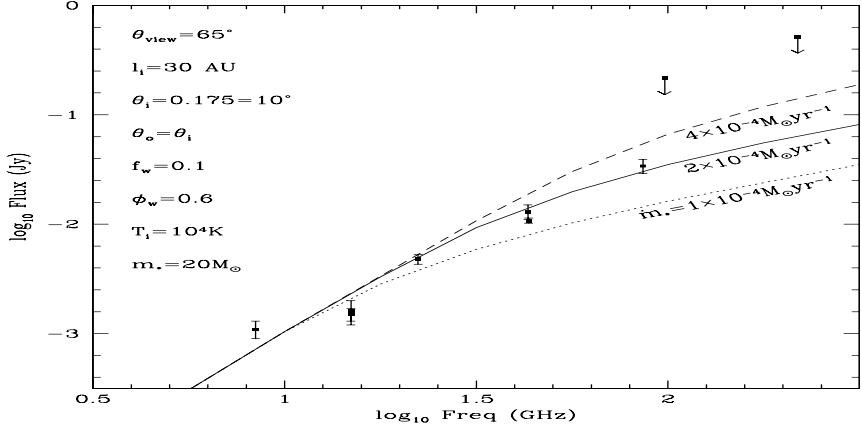}{2in}{0}{60}{30}{-180}{-50}
\label{fig:jtan2}
\caption{Radio continuum spectrum of source ``I'' in the Orion hot
core. Data at $\leq$86~GHz are tabulated in Plambeck et
al. (1995). Additional data and limits at 43.1, 98, and 218~GHz are
from Chandler \& Wood (1997), Murata et al. (1992) and Blake et
al. (1996), respectively. Models are described in the text.}\vspace{-0.2in}
\end{figure}

At a distance of 450~pc, the Orion hot core is the closest example of
a massive star in formation. The core is self-luminous ($L_{\rm
bol}\sim 1-5\times 10^4\:{\rm L_\odot}$, Gezari et al. 1998 [G98]; Kaufman et al. 1998). A
weak radio continuum source (``I'') (e.g. Menten \& Reid 1995),
located within a few arcseconds of the core center, as traced by dust
and gas emission (Wright, Plambeck \& Wilner 1996),
almost certainly pinpoints the location of the
massive protostar. Note that the Becklin-Neugebauer (BN) object
($L\simeq 10^4\:{\rm L_\odot}$; 0.02~pc to the NW in projection) is
likely to be a runaway B star (Plambeck et al. 1995; Plambeck 2002,
private comm.), that did not form close to the hot core; radio source
``L'' (0.007~pc to the SW in projection; sometimes referred to as
``n'') is not particularly embedded or luminous (G98) and may be an
intermediate mass protostar. Note also that bright near and mid-infrared features
(e.g. IrC2) often result from inhomogeneous extinction (G98). X-ray observations (Garmire et
al. 2000) can detect lower-mass protostars to $A_V\sim 60$ and reveal several sources (including ``L'')
within about 0.02~pc projected distance of source ``I'', which itself
is not detected.

A large scale, wide-angle bipolar outflow extends to the NW and SE of
the core (Chernin \& Wright 1996, and references therein). These
authors have modeled the flow as being inclined at $65^\circ$ to our
line of sight. At 22~GHz, source ``I'' appears elongated (0\arcsec.145
by $<0\arcsec.085$, Menten 2002, private comm.) parallel to the large
scale outflow axis. In a perpendicular direction, SiO emission forms a
``bow-tie'' feature centered on ``I'', that may be an inclined
or flared disk (Wright et al. 1995). SiO should be abundant where dust
grains are destroyed.

We model the thermal radio emission from source ``I'' assuming the
ionized gas has a density given by equation (\ref{nangular}) for two
jets of fixed half-opening angle $\theta_i=10^\circ$ and length
$l_i=30$~AU, inclined at $65^\circ$ to our line of sight. These
dimensions are consistent with the 22~GHz size of $0\arcsec.145$ and the lower frequency fluxes being due to optically thick
emission at temperature $T_i=10^4\:{\rm K}$. We investigate models
with $\theta_0=\theta_i$ and $\dot{m}_*=1,2,4\times
10^{-4}M_\odot {\rm yr^{-1}}$. Fig.~2 lists other parameters.  A consistent,
but not unique, model accounts for the radio spectrum with an
outflow from a $20M_\odot$ protostar accreting at
$\dot{m}_*\simeq 2\times 10^{-4}M_\odot {\rm yr^{-1}}$.  
More realistic geometries of the ionized region, derived from the predicted ionizing flux,
are considered in a future paper (Tan \& McKee 2002, in prep.).

\section{Conclusions}

The collisional model for star formation can account qualitatively for
some of the observational properties of massive stars, such as their
tendency to form in the centers of clusters. However, it is difficult
to achieve the necessary stellar densities ($\sim 10^8\:{\rm
pc^{-3}}$) for this process to be efficient. The standard accretion
model, modified to account for the high pressures and turbulent,
nonthermal conditions of massive star-forming clumps can achieve the
high-accretion rates necessary to overcome radiation pressure and
achieve short formation timescales. Crowding is not a serious problem.
Nearby massive star-forming regions show signatures of disks and
collimated outflows, suggesting the accretion picture is relevant to
the formation of stars with masses up to at least $\sim 20-30\sm$.

\acknowledgments This article
summarizes work carried out in collaboration with Christopher
F. McKee. I thank Dick Plambeck and Karl Menten for sharing
unpublished results on Orion.  My research is supported by a
Spitzer-Cotsen fellowship from Princeton University.

\end{document}